\documentclass[english,english,aps,prl,graphicx,twocolumn]{revtex4}
\usepackage{graphicx}
\usepackage{subfigure}
\makeatletter

\newcommand{\br}{{\bf r}}
\newcommand{\bx}{{\bf x}}
\usepackage{babel}

\begin{document}

\title{Failure of the random phase approximation correlation energy}

\author{Paula Mori-S\'anchez}
\author{Aron Cohen}
\altaffiliation{Department of Chemistry, University of Cambridge, Lensfield Road, Cambridge, CB2 1EW, UK}
\author{Weitao Yang}

\affiliation{Department of Chemistry, Duke University, Durham, North Carolina 27708, USA}

\date{\today{}}

\begin{abstract}
The random phase approximation (RPA) to the correlation energy is extended to
fractional occupations and its performance examined for exact conditions on fractional charges
and fractional spins. RPA satisfies the constancy condition for fractional spins that
leads to correct bond dissociation and no static correlation error for H$_2$ but massively
fails for fractional charges, with an enormous delocalization error even for a
one-electron system such as H$_2^+$. Other methods such as range-separated RPA 
can reduce this delocalization error but only at the cost of increasing the static
correlation error.  None of the RPA methods seem to have the discontinuous nature
required to satisfy both exact conditions and the full unified condition, emphasizing the need to 
go further than just smooth functionals of the orbitals.
\end{abstract}
\maketitle

The Random Phase Approximation (RPA) \cite{RPA1}, formulated
within the adiabatic-connection fluctuation dissipation theorem, provides
an appealing definition of the exchange-correlation energy. It can also
be simply viewed \cite{DFTRPA} from a density-functional theory (DFT) perspective as
a functional of all (occupied plus virtual) orbitals and eigenvalues
in the Kohn-Sham (KS) formalism.  There has been much recent attention
to the RPA \cite{FurcheRPA08,ScuseriaRPAlr,SavinRPAlr,KresseRPA,RubioRPAgap,BurkeRPAh2}, highlighted by the work of Furche \cite{FurcheRPA08},
who has shown a practical way to calculate the
correlation energy in the KS context in a similar fashion as in wave function approach \cite{bookallrpa}. 
Much of the interest in RPA comes from the improved description of two
key aspects  for which many other density-functional approximations
(DFAs) encounter several problems:  One is the description of weak  and
Van-der-Waals interactions, as typified by molecules such as He$_2$ or
Ne$_2$; the other is the description of static correlation as seen in
the stretching of H$_2$ \cite{BurkeRPAh2} and N$_2$ \cite{FurcheRPA01}.
Other efforts to include unoccupied orbitals into the exchange-correlation energy, such as MP2 or GL2, have
not been so successful with unphysical divergence for very simple systems
\cite{OEPMP2}. Also the idea of range separation has been applied to
the RPA correlation \cite{SavinRPAlr}, and interesting functionals including
long-range RPA have been developed \cite{ScuseriaRPAlr}.

We have recently been attempting to gain insight \cite{Science} into the problems of DFAs and
other electron correlation methods relating to the delocalization error
\cite{Delocalization} (fractional charge perspective) and the static
correlation error \cite{Fracspin} (fractional spin perspective). The
combination of these two in a unified condition \cite{Flatplane} is much
more powerful and highlights  a basic property of the exchange-correlation
functional, that is its discontinuous derivative at integer number of
particles in particular for fractional spin systems. This property is violated by all functionals in the literature
and it is key, for example, for the accurate calculation  the the band-gap
of strongly correlated systems. The performance of a method can be
investigated by testing this unified condition for fractional charges
and spins, which just requires a generalization of the method
to include occupation numbers and fractional occupations as we have done recently
for MP2 \cite{MP2frac}.
 This test is remarkably simple and illuminating,
as it is sufficient to perform  calculations on one-electron systems (e.g. the
Hydrogen atom even with one basis function is enough).  Our aim in
this work is to carry out this simple test for RPA and investigate  some
of the implications for the method. This understanding offers insight
about the largest errors of DFAs,
but does not address the weak but very important Van-der-Waals
energies.

Consider the matrix representation of the RPA problem  \cite{FurcheRPA01}
\begin{equation}
\left( \begin{array}{cc} {\bf A} &{\bf B} \\ {\bf -B} &{\bf -A} \end{array} \right)
\left( \begin{array}{c} {\bf X} \\ {\bf Y}  \end{array} \right)
=
\omega
\left( \begin{array}{c} {\bf X} \\ {\bf Y},  \end{array} \right)
\label{rpaeigen}
\end{equation}
where the matrices ${\bf A, B, X, Y}$ are of dimension $n_{occ}n_{virt}\times n_{occ}n_{virt}$, with $n_{occ}$ and $n_{virt}$ being the number of occupied and virtual orbitals, and 
${\bf \omega}$ is the $n_{occ}n_{virt}$  vector of excitation energies.
RPA is given by the solution of the above  equations in the KS orbital basis with 
\begin{eqnarray}
\label{aba}
A_{ia,jb} &=& (\epsilon_a - \epsilon_i)\delta_{ia,jb} + \langle ij|ab\rangle  \\
B_{ia,jb} &=& \langle ij|ab\rangle 
\label{abb}
\end{eqnarray}
where $\epsilon$ are KS eigenvalues, $i,j$ are occupied orbitals, $a,b$ are virtual orbitals,
and $\langle ij|ab\rangle  = \int\int \frac{\phi_i(\bx)\phi_a(\bx)\phi_j(\bx^\prime)\phi_b(\bx^\prime)}{|\br-\br^\prime|} {\rm d}\bx {\rm d}\bx^\prime $
where $\bx$ is a combined space and spin coordinate.  This corresponds to
a Hartree-only density response with no exchange-correlation contribution.
RPAE (also called RPA+X or full RPA) includes a Hartree-Fock response that requires
antisymmetrized $\langle ij||ab\rangle =\langle ij|ab\rangle -\langle ij|ba\rangle $ 
integrals in Eqs. \ref{aba},\ref{abb}.

To extend the 
method to fractional occupation of the orbitals, the 
occupation numbers $\{n_p\}$ can be included into the basic matrices  using an extension of the fluctuation-dissipation theorem to ensemble density matrices \cite{rpaweitao}
\begin{eqnarray}
\nonumber
A_{ia,jb} &=& (\epsilon_a - \epsilon_i)\delta_{ia,jb} \\
          & &  + \langle ij|ab\rangle  \sqrt{n_in_j(1-n_a)(1-n_b)} \\
B_{ia,jb} &=& \langle ij|ab\rangle  \sqrt{n_in_j(1-n_a)(1-n_b)}. 
\label{rpafrac}
\end{eqnarray}
Partially occupied orbitals are considered both occupied and
virtual, such that now $i,j$ run over $n_{occ}+n_{frac}$, $a,b$ run over $n_{frac}+n_{virt}$ 
and the dimensionality of the matrices extends to
$(n_{occ}+n_{frac})(n_{virt}+n_{frac})\times(n_{occ}+n_{frac})(n_{virt}+n_{frac})$
for $n_{frac}$ number or fractionally occupied orbitals.
This is consistent with the perspective of fractional charges and spins resulting from dissociation. At the dissociation limit the HOMO and LUMO both become fractional. This appears to be a correct recipe to extend functionals of orbitals to fractional occupations. It gives a correct extension for all the orbital functionals 
with  occupied orbitals (GGA, HF) and also for MP2. 
The RPA correlation energy is given by \cite{FurcheRPA08}
\begin{equation}
E_c^{\rm RPA} = \frac{1}{2}\sum_{ia} (\omega_{ia} - A_{ia,ia})
\end{equation}
with no additional (frequency or coupling constant) integrations and fully expressed in term of
KS quantities.
It should be noted that it is also possible to calculate  the derivatives 
$\frac{\partial E}{\partial N}$ 
by taking derivatives with respect to the frontier orbital occupation number, 
$\frac{\partial E}{\partial N} = \frac{\partial E_c^{\rm RPA}}{\partial n_f}$,
to get the band-gap as previously  done for MP2 \cite{MP2frac}.

\begin{figure}[!t]
\includegraphics[width=0.5\textwidth]{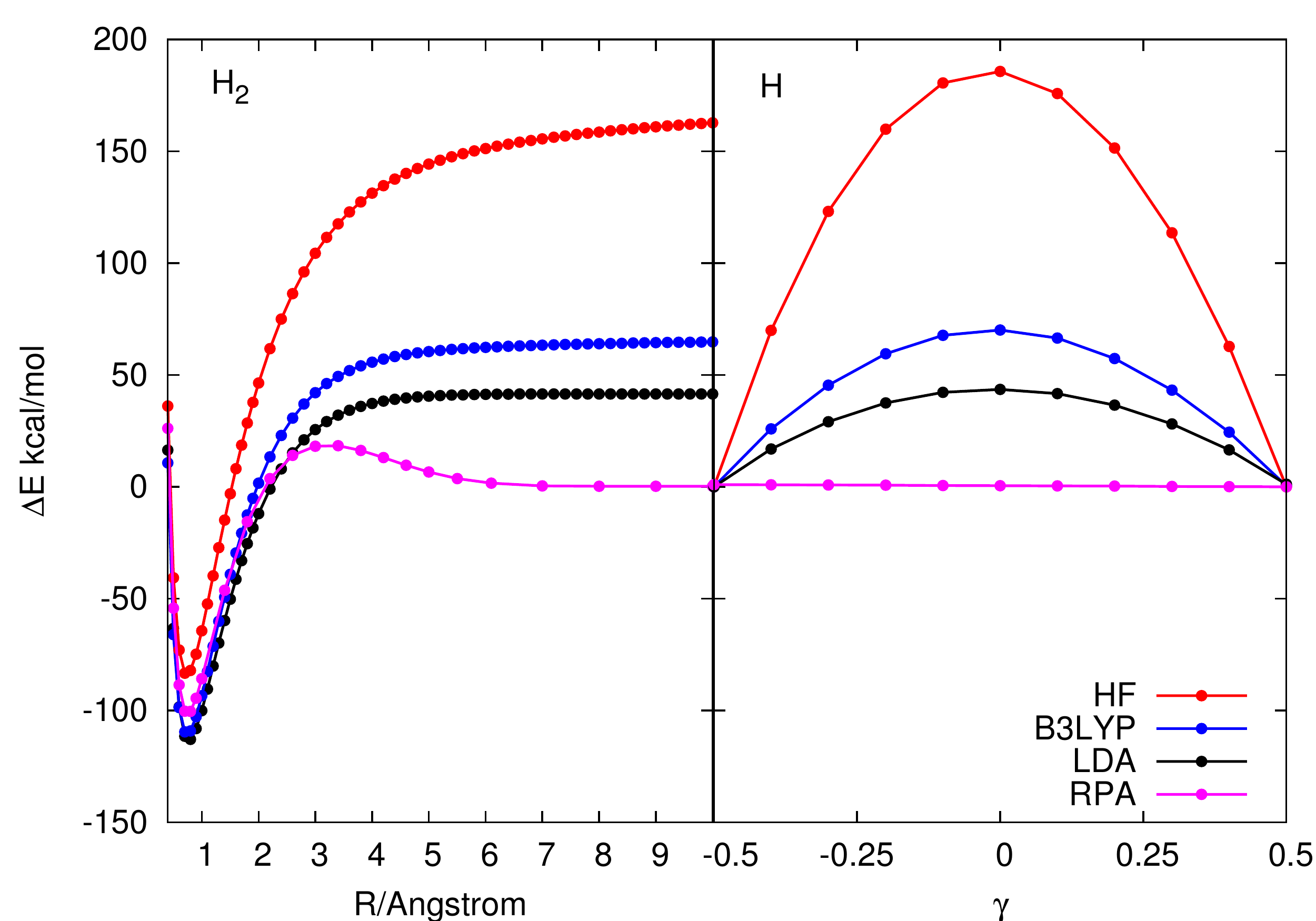}
\caption{The closed-shell dissociation of H$_2$ compared with the fractional spin H atom
$n_\alpha = \frac{1}{2} + \gamma, n_\beta = \frac{1}{2} - \gamma$. All calculations,
except Fig. \ref{flatplanelr}, use a cc-pVQZ basis set.}
\label{h2}
\end{figure}

We have implemented the above equations in a modified version of CADPAC. We do this very simply by
calculating the whole $\bf A$ and $\bf B$ matrices and then diagonalizing
according to Eq. \ref{rpaeigen} to give the excitation energies.  This is computed
post-PBE calculation using the fractional PBE KS orbitals
and eigenvalues, to give the total RPA exchange-correlation energy
$E_{xc}= E_x^{\rm EXX}+ E_c^{\rm RPA}$. 
This functional could be also treated
in a variational fashion using the optimized effective potential method
(or its generalized version to accommodate nonlocal potentials in the
case of for RPAE), but this has not been done in this work.

\begin{figure}[!t]
\includegraphics[width=0.5\textwidth]{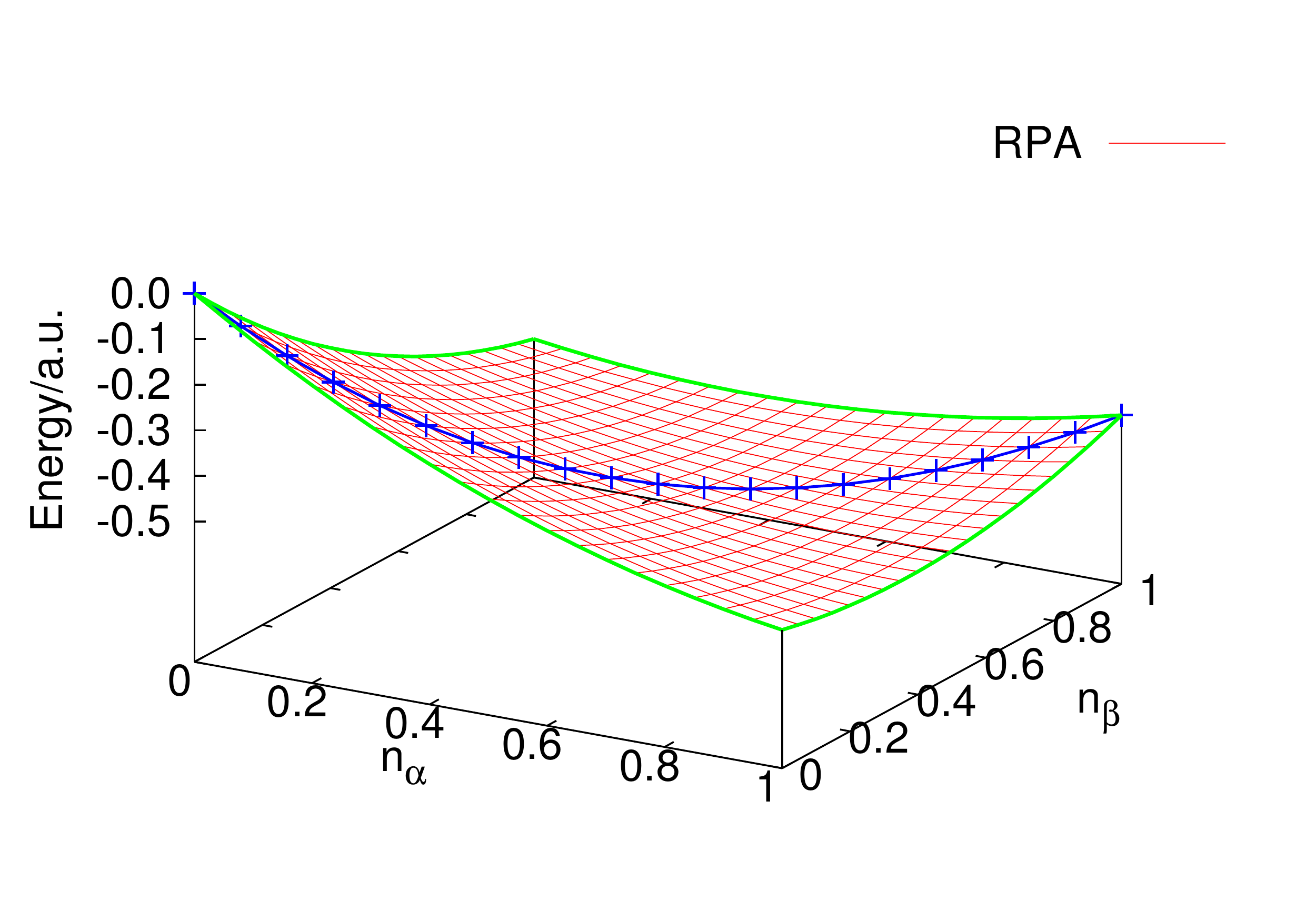}
\caption{The energy of H$[n_\alpha,n_\beta]$ for $0\le n_\alpha \le 1$ and $0\le n_\beta \le 1$.}
\label{rpaflatplane}
\end{figure}

\begin{figure}[!b]
\includegraphics[width=0.5\textwidth]{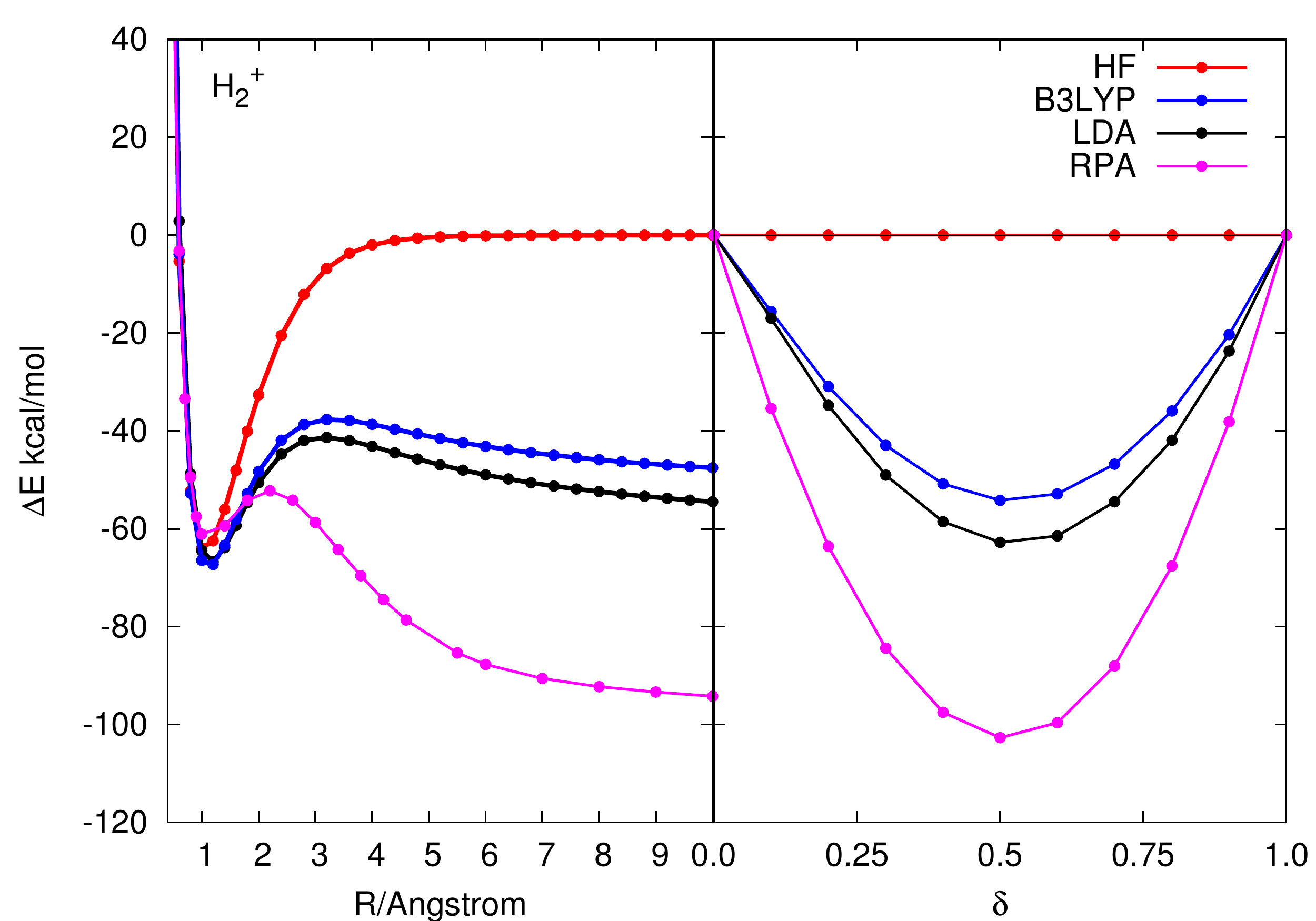}
\caption{The dissociation of H$_2^+$ compared with the fractional charge H atom,
$n_\alpha = \delta, n_\beta=0$.}
\label{h2plus}
\end{figure}


One of the promising aspects of RPA is that it greatly improves upon
DFAs for the closed shell dissociation of H$_2$. This is illustrated in
Fig. \ref{h2}, where the RPA energy is compared with LDA, Hartree-Fock (HF)
and B3LYP \cite{B3LYP}. RPA predicts the correct dissociation limit, which clearly
correlates with the much improved behavior of the Hydrogen atom with
fractional spins, H$[\frac{1}{2},\frac{1}{2}]$. Remarkably, RPA satisfies the constancy
condition, and all fractional spin configurations are degenerate in
energy and equal to that of the pure spin H atom with $[1,0]$. It is
possible to study in 
more detail the  energy of the Hydrogen atom with general spin
occupations, $E[n_\alpha,n_\beta]$, as is shown in Fig. \ref{rpaflatplane}. The exact energy
should be two flat planes that intersect with a line of discontinuity at
$n_\alpha+n_\beta=1$. We have shown previously \cite{Flatplane} that smooth functionals of
the occupied orbitals, such as LDA, GGA, HF and other hybrid functionals,
are unable to qualitatively give this discontinuous behavior of the
$E[n_\alpha,n_\beta]$ surface. Other methods involving virtual orbitals
such as MP2 or its degenerate corrected version also fail \cite{MP2frac}.  For RPA, it
might not be immediately clear from a consideration of the underlying
equations whether  the RPA energy  has a discontinuity that is
sufficient to give the desirable flat plane  behavior.  A simple test
on the Hydrogen atom shows that RPA also qualitatively fails and misses
the discontinuity. Therefore, it is expected to fail for problems where
this discontinuity is key, such as the band-gap  of strongly-correlated
systems.

\begin{figure}[!t]
\includegraphics[width=0.5\textwidth]{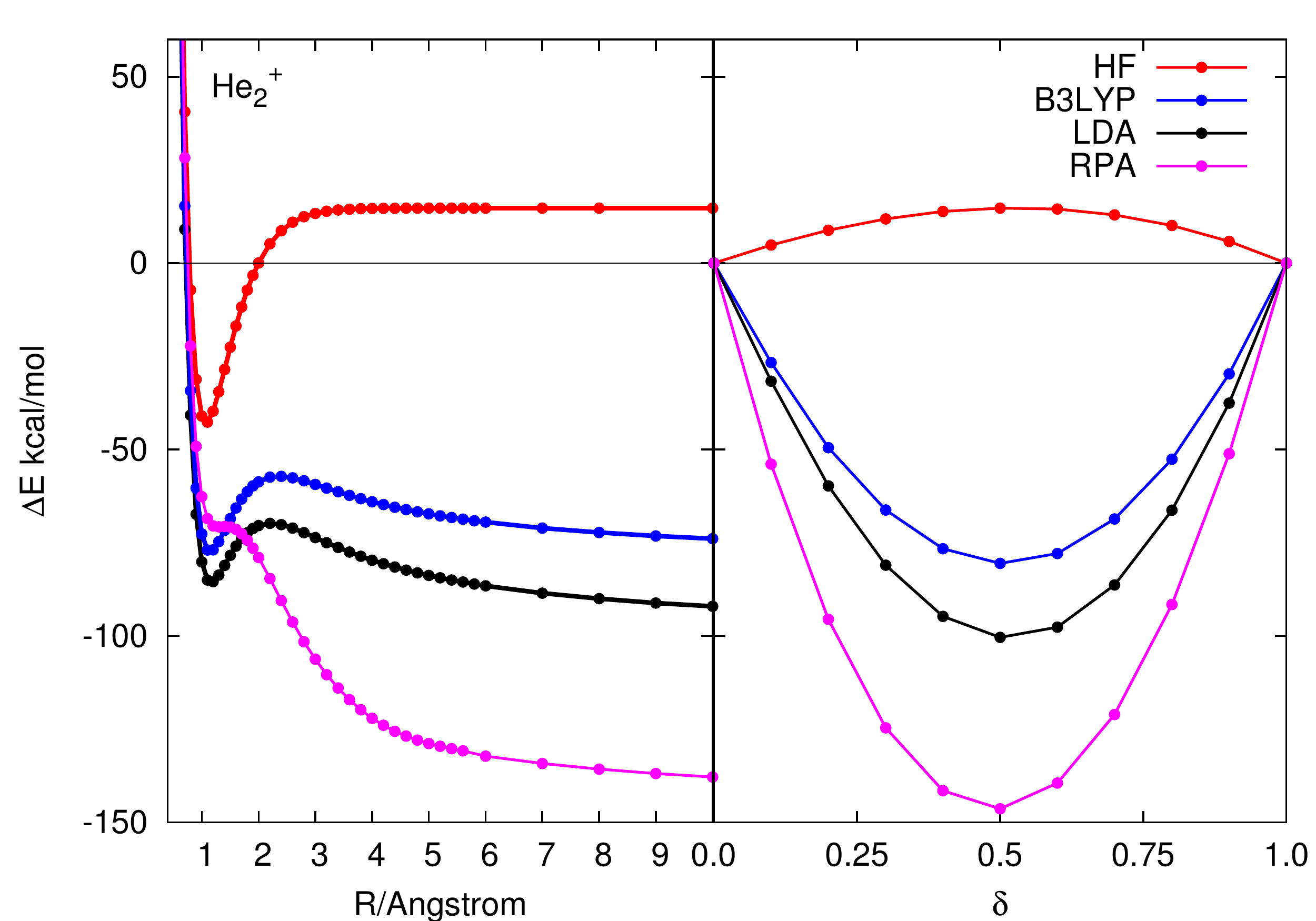}
\caption{The dissociation of He$_2^+$ compared with the fractional
charge He atom, 
$n_{\alpha,1}=1,
n_{\alpha,2}=1;
\ \ 
n_{\beta,1}=1,
n_{\beta,2}=\delta$.}
\label{he2plus}
\end{figure}

\begin{figure}[!b]
\includegraphics[width=0.5\textwidth]{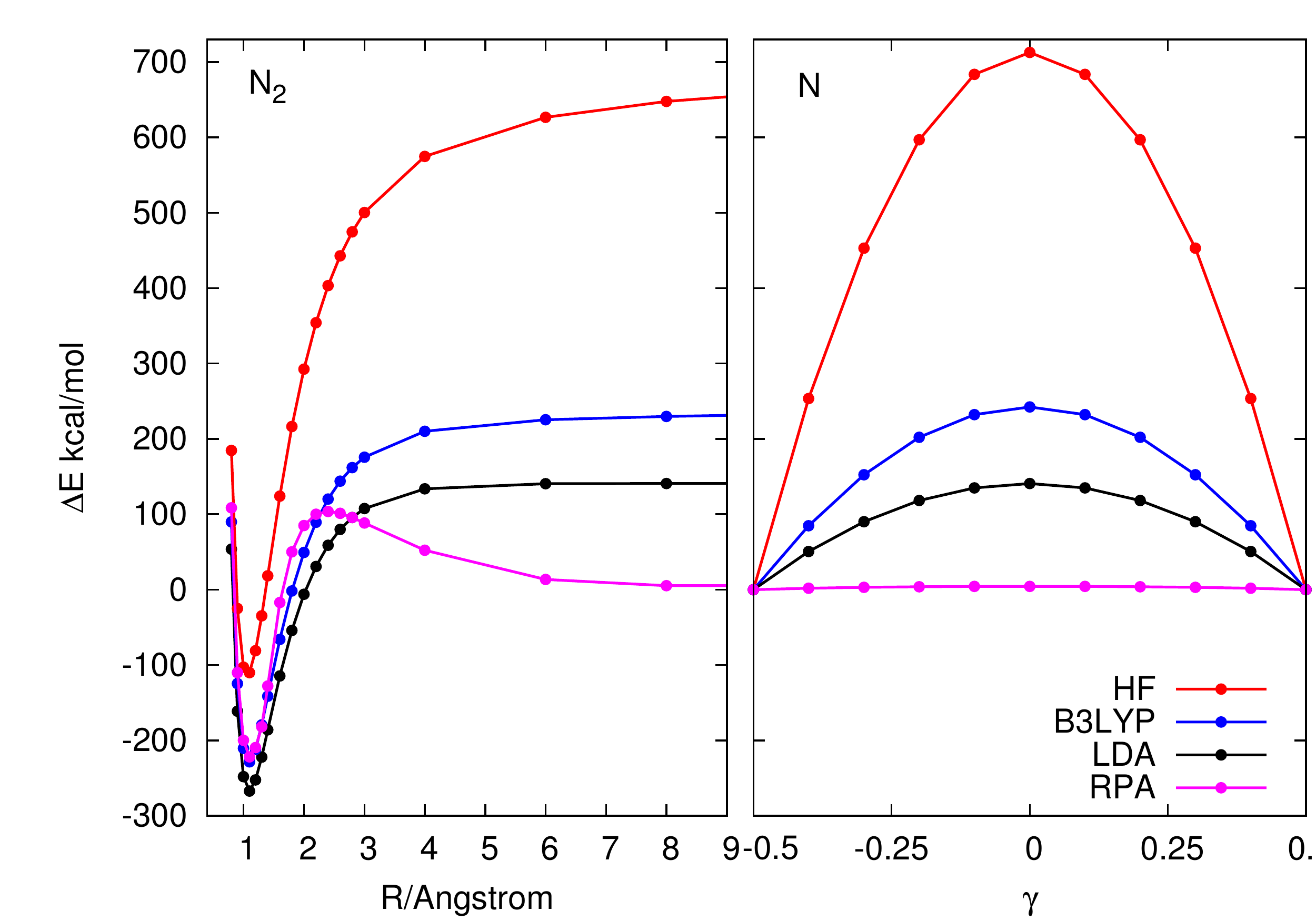}
\caption{The closed shell dissociation of N$_2$ compared with the fractional spin
N atom,
$n_{\alpha,1}=1,
n_{\alpha,2}=1,
n_{\alpha,3}=n_{\alpha,4}=n_{\alpha,5}=(\frac{1}{2}+\gamma);
\ \ \ \ 
n_{\beta,1}=1,
n_{\beta,2}=1,
n_{\beta,3}=n_{\beta,4}=n_{\beta,5}=(\frac{1}{2}-\gamma)$.}
\label{n2}
\end{figure}

\begin{figure}[!t]
\includegraphics[width=0.5\textwidth]{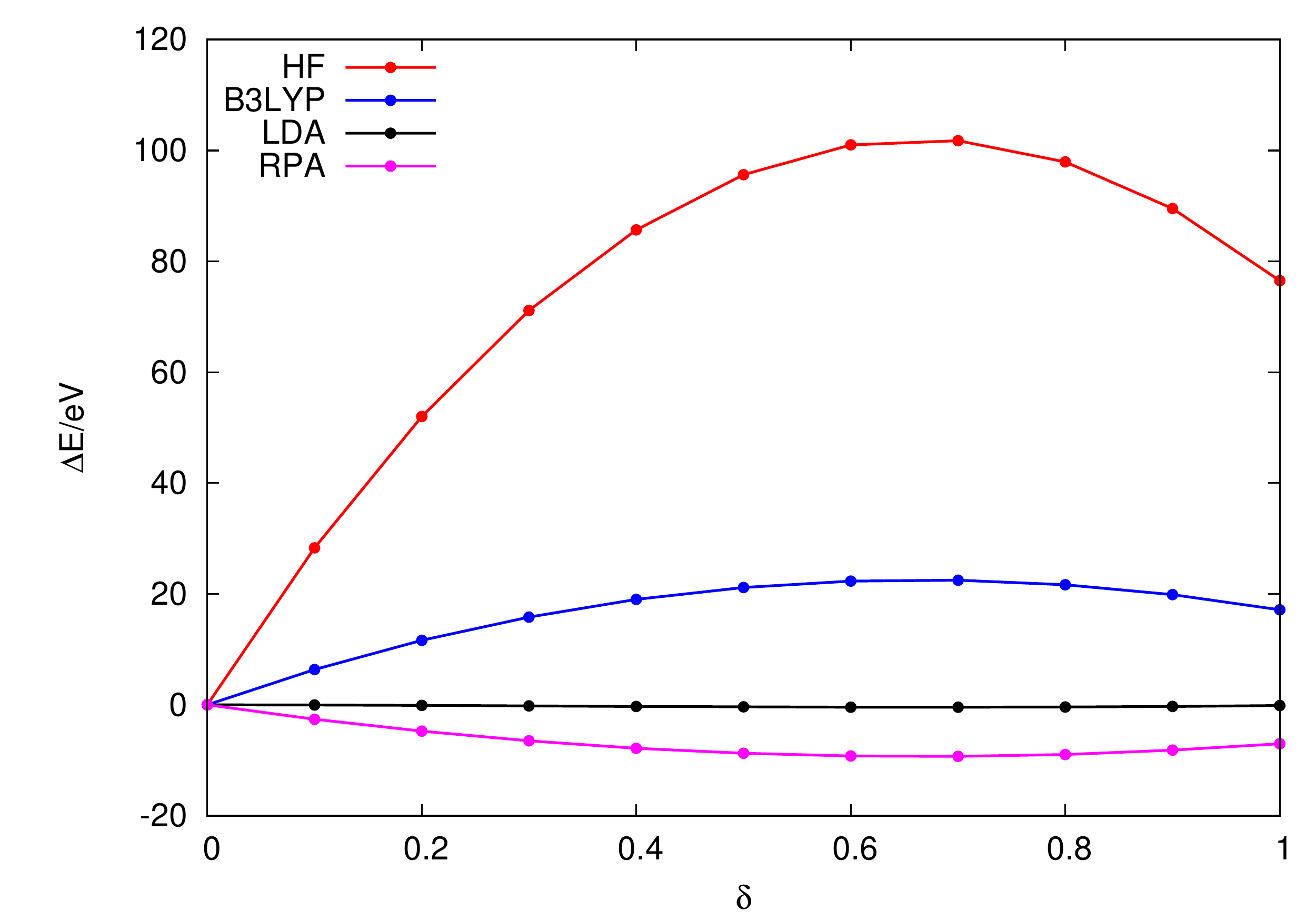}
\caption{A fractional spin Carbon atom calculated with 
$n_{\alpha,1}=1,
n_{\alpha,2}=1,
n_{\alpha,3}=n_{\alpha,4}=\left(1-\frac{\delta}{2}\right),n_{\alpha,5}=\delta;
\ \ \ \ 
n_{\beta,1}=1,
n_{\beta,2}=1$. The spherical Carbon atom corresponds to $\delta=\frac{2}{3}$.
}
\label{spherical}
\end{figure}

Very importantly, Fig. \ref{rpaflatplane} reveals another real problem of RPA for
the treatment of fractional charges, with an extremely convex behavior,
much more so than LDA  or other DFAs.  To emphasize this erroneous
behavior, Fig. \ref{h2plus}  shows the dissociation curve
of stretched H$_2^+$.  The result is astonishing, a simple one-electron
system where RPA behaves extremely badly and give massive correlation energies. 
Another simple and paradigmatic
case, the dissociation of He$_2^+$ is considered in Fig. \ref{he2plus}. Again RPA
fails dramatically, as shown for the He with
fractional charges,  leading to unphysically low correlation energies
that affect not only the dissociation but also the bonding region.
Again this error may not be obvious from the underlying equation, but
is revealed by extremely simple tests which highlight many important
problems of the method. It is now clear that RPA suffers from large
delocalization error, which might be  due
to the lack of an underlying wavefunction and the poor quality
of the Hartree-only response in RPA.  This error is
pervasive and can be seen in calculations of many different systems and
properties.  Thus the positive aspect of RPA in the improved description
of Van-der-Waals systems such as He$_2$  gets clouded by the spectacular
failure to describe related systems such as He$_2^+$.

On a more positive note, we examine the behavior of RPA for some other
challenging static correlation error problems. In general, the same
improvement seen for H$_2$ carries over to more complex systems, such as
N$_2$ in Fig. \ref{n2}, where RPA is still very close to satisfying the
constancy condition for multiple fractional spins.  Fig. \ref{spherical}
considers the slightly different static correlation problem of the spatial
degeneracy within the $p$ set, as exemplified by the massive error of
HF for an spherical atom against a non-spherical atom.  In this case
RPA makes a large correction to HF, and rather interestingly, it  even
makes the spherical atom sightly lower in energy, in contrast to most
methods. These two positve aspects show that RPA deals well with this type
of spin and spatial degeneracy and makes it a very interesting prospect
to tackle more challenging problems of transtition metal chemistry.

\begin{figure}[!t]
\includegraphics[width=0.5\textwidth]{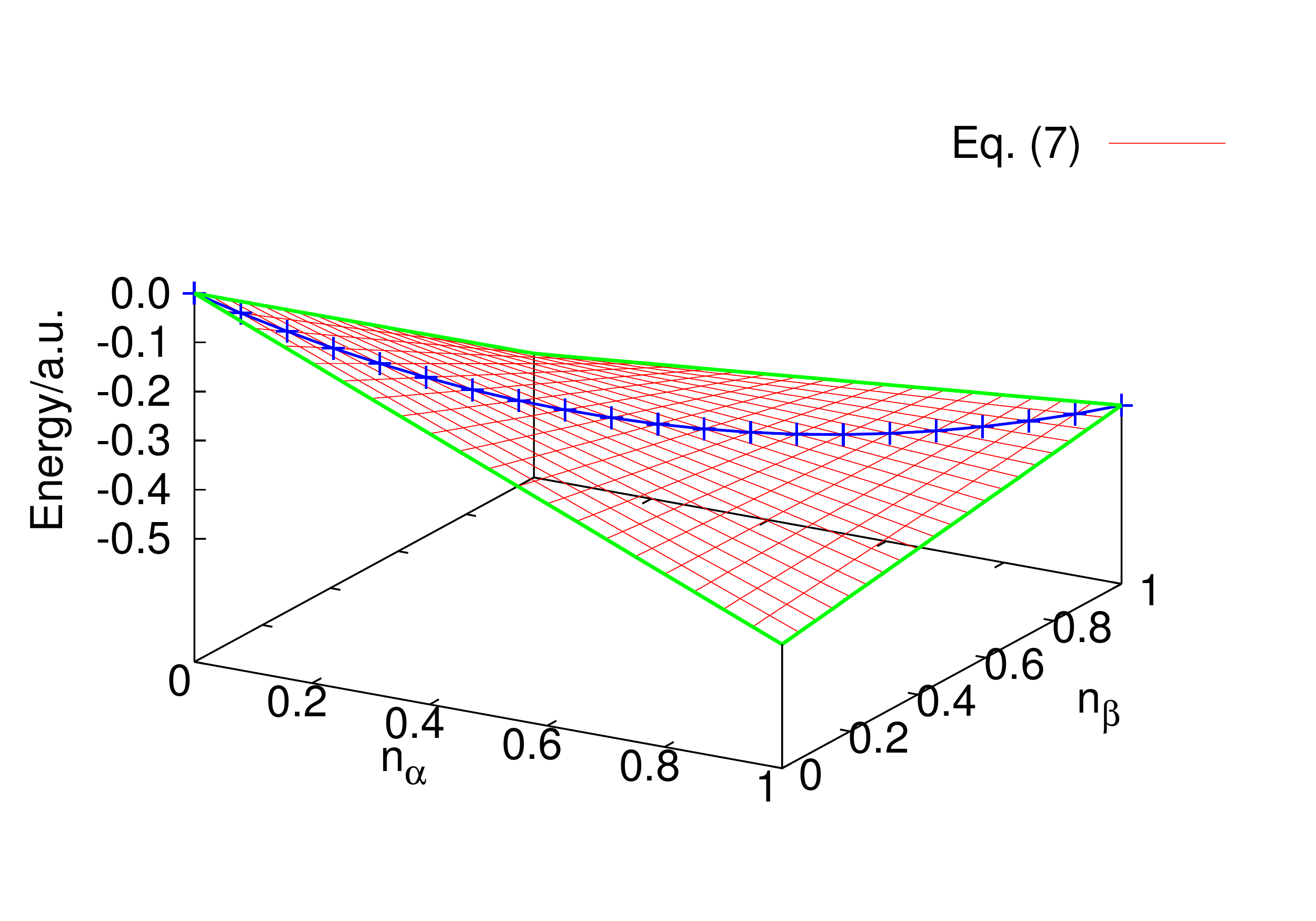}
\caption{The same as Fig. (\ref{rpaflatplane}) except for a range-separated RPA functional, Eq. (\ref{rpalr}), calculated with a 6-31G basis set. }
\label{flatplanelr}
\end{figure}

This analysis should be applied to any method
to gain a deeper insight into its behavior.  For example
RPAE may not suffer from a massive delocalization error but correspondingly it would no longer satisfy the 
constancy condition, performing worse
for fractional spins and hence H$_2$ dissociation. It is extremely difficult to 
improve both aspects at the same time and in a manner that leads to the flat plane behavior.
There has also been much recent interest in including range-separation
into the RPA ingredients. 
Following the work of Janesko {\it et al.}
\cite{ScuseriaRPAlr} we examine in Fig. \ref{flatplanelr} 
\begin{equation}
E_{xc} = E_x^{\rm SR,LDA} + E_x^{\rm LR,HF} + E_c^{\rm VWN} + E_c^{\rm LR,RPA}
\label{rpalr}
\end{equation}
where a value of $\mu=1.2$ a.u. is used for the range separation parameter and the $E_x^{\rm SR,LDA}$ 
is from Iikura {\it et al.} \cite{HiraoLR}. The $E_c^{\rm LR,RPA}$ is RPA with the long-range integrals in Eqs. \ref{aba},\ref{abb}.
This energy is evaluated using PBE orbitals and eigenvalues.
Fig. \ref{flatplanelr} clearly illustrates that the effect of range separation is in general
to move the surface up such that the error for fractional charges  is
decreased, however  the good performance for fractional spin deteriorates
correspondingly and the constancy condition is no longer fullfilled.
There are many other possibilities in methods related to RPA and the
fluctuation-dissipation theorem, such as changing the eigenvalues
(KS versus generalized KS) and the kernel (going from RPA to
RPAE). We would recommend that while developing such methods 
at least stretched H$_2^+$ and H$_2$ are considered, and if the extension
to fractional occupations is possible the flat-plane behavior of
Fig. \ref{rpaflatplane} is investigated.  


We have investigated the behavior of some RPA based methods for the
extension to fractional charges and fractional spins, to give insight into
their performance for harder challenges in theoretical chemistry. Although
RPA correlation is a complicated orbital dependent functional,
it still does not have the discontinuous behavior needed to satisfy both
the exact conditions for fractional charge and
fractional spin. RPA has a much reduced static correlation error
but has a massive delocalization error.  In fact, the lack of discontinuous
nature means that an improved performance for fractional spins likely leads
to a much worse performance for fractional charges, and  vice versa.
For example with range separation the fractional charge
behavior improves but the dissociation limit of H$_2$ worsens. The
main purpose of this work is to highlight the importance of the exact
conditions on the energy to be satisfied by methods in and outside DFT,
and the complexity that is needed may not be covered easily by functionals
of the unoccupied orbitals and eigenvalues.

Support from National Science Foundation and the Royal Society (AJC)  is greatly appreciated.


\begin{thebibliography}{19}
\expandafter\ifx\csname natexlab\endcsname\relax\def\natexlab#1{#1}\fi
\expandafter\ifx\csname bibnamefont\endcsname\relax
  \def\bibnamefont#1{#1}\fi
\expandafter\ifx\csname bibfnamefont\endcsname\relax
  \def\bibfnamefont#1{#1}\fi
\expandafter\ifx\csname citenamefont\endcsname\relax
  \def\citenamefont#1{#1}\fi
\expandafter\ifx\csname url\endcsname\relax
  \def\url#1{\texttt{#1}}\fi
\expandafter\ifx\csname urlprefix\endcsname\relax\def\urlprefix{URL }\fi
\providecommand{\bibinfo}[2]{#2}
\providecommand{\eprint}[2][]{\url{#2}}

\bibitem[{\citenamefont{Bohm and Pines}(1952)}]{RPA1}
\bibinfo{author}{\bibfnamefont{D.}~\bibnamefont{Bohm}} \bibnamefont{and}
  \bibinfo{author}{\bibfnamefont{D.}~\bibnamefont{Pines}},
  \bibinfo{journal}{Phys. Rev.} \textbf{\bibinfo{volume}{85}},
  \bibinfo{pages}{338} (\bibinfo{year}{1952}).

\bibitem[{\citenamefont{Langreth and Perdew}(1975)}]{DFTRPA}
\bibinfo{author}{\bibfnamefont{D.~C.} \bibnamefont{Langreth}} \bibnamefont{and}
  \bibinfo{author}{\bibfnamefont{J.~P.} \bibnamefont{Perdew}},
  \bibinfo{journal}{Solid State Comm.} \textbf{\bibinfo{volume}{17}},
  \bibinfo{pages}{1425} (\bibinfo{year}{1975}).

\bibitem[{\citenamefont{Furche}(2008)}]{FurcheRPA08}
\bibinfo{author}{\bibfnamefont{F.}~\bibnamefont{Furche}}, \bibinfo{journal}{J.
  Chem. Phys.} \textbf{\bibinfo{volume}{129}}, \bibinfo{pages}{114105}
  (\bibinfo{year}{2008}).

\bibitem[{\citenamefont{Janesko et~al.}(2009)\citenamefont{Janesko, Henderson,
  and Scuseria}}]{ScuseriaRPAlr}
\bibinfo{author}{\bibfnamefont{B.~G.} \bibnamefont{Janesko}},
  \bibinfo{author}{\bibfnamefont{T.~M.} \bibnamefont{Henderson}},
  \bibnamefont{and} \bibinfo{author}{\bibfnamefont{G.~E.}
  \bibnamefont{Scuseria}}, \bibinfo{journal}{J. Chem. Phys.}
  \textbf{\bibinfo{volume}{130}}, \bibinfo{pages}{081105}
  (\bibinfo{year}{2009}).

\bibitem[{\citenamefont{Toulouse et~al.}(2009)\citenamefont{Toulouse, Gerber,
  Jansen, Savin, and \'Angy\'an}}]{SavinRPAlr}
\bibinfo{author}{\bibfnamefont{J.}~\bibnamefont{Toulouse}},
  \bibinfo{author}{\bibfnamefont{I.~C.} \bibnamefont{Gerber}},
  \bibinfo{author}{\bibfnamefont{G.}~\bibnamefont{Jansen}},
  \bibinfo{author}{\bibfnamefont{A.}~\bibnamefont{Savin}}, \bibnamefont{and}
  \bibinfo{author}{\bibfnamefont{J.~G.} \bibnamefont{\'Angy\'an}},
  \bibinfo{journal}{Phys. Rev. Lett.} \textbf{\bibinfo{volume}{102}},
  \bibinfo{pages}{096404} (\bibinfo{year}{2009}).

\bibitem[{\citenamefont{Harl and Kresse}(2008)}]{KresseRPA}
\bibinfo{author}{\bibfnamefont{J.}~\bibnamefont{Harl}} \bibnamefont{and}
  \bibinfo{author}{\bibfnamefont{G.}~\bibnamefont{Kresse}},
  \bibinfo{journal}{Phys. Rev. B} \textbf{\bibinfo{volume}{77}},
  \bibinfo{pages}{045136} (\bibinfo{year}{2008}).

\bibitem[{\citenamefont{Gruning et~al.}(2006)\citenamefont{Gruning, Marini, and
  Rubio}}]{RubioRPAgap}
\bibinfo{author}{\bibfnamefont{M.}~\bibnamefont{Gruning}},
  \bibinfo{author}{\bibfnamefont{A.}~\bibnamefont{Marini}}, \bibnamefont{and}
  \bibinfo{author}{\bibfnamefont{A.}~\bibnamefont{Rubio}}, \bibinfo{journal}{J.
  Chem. Phys.} \textbf{\bibinfo{volume}{124}}, \bibinfo{pages}{154108}
  (\bibinfo{year}{2006}).

\bibitem[{\citenamefont{Fuchs et~al.}(2005)\citenamefont{Fuchs, Niquet, Gonze,
  and Burke}}]{BurkeRPAh2}
\bibinfo{author}{\bibfnamefont{M.}~\bibnamefont{Fuchs}},
  \bibinfo{author}{\bibfnamefont{Y.~M.} \bibnamefont{Niquet}},
  \bibinfo{author}{\bibfnamefont{X.}~\bibnamefont{Gonze}}, \bibnamefont{and}
  \bibinfo{author}{\bibfnamefont{K.}~\bibnamefont{Burke}}, \bibinfo{journal}{J.
  Chem. Phys.} \textbf{\bibinfo{volume}{122}}, \bibinfo{pages}{094116}
  (\bibinfo{year}{2005}).

\bibitem[{\citenamefont{Ring and Schuck}(2004)}]{bookallrpa}
\bibinfo{author}{\bibfnamefont{P.}~\bibnamefont{Ring}} \bibnamefont{and}
  \bibinfo{author}{\bibfnamefont{P.}~\bibnamefont{Schuck}},
  \emph{\bibinfo{title}{The nuclear many-body problem}}
  (\bibinfo{publisher}{Springer}, \bibinfo{year}{2004}).

\bibitem[{\citenamefont{Furche}(2001)}]{FurcheRPA01}
\bibinfo{author}{\bibfnamefont{F.}~\bibnamefont{Furche}},
  \bibinfo{journal}{Phys. Rev. B} \textbf{\bibinfo{volume}{64}},
  \bibinfo{pages}{195120} (\bibinfo{year}{2001}).

\bibitem[{\citenamefont{Mori-Sanchez et~al.}(2005)\citenamefont{Mori-Sanchez,
  Wu, and Yang}}]{OEPMP2}
\bibinfo{author}{\bibfnamefont{P.}~\bibnamefont{Mori-Sanchez}},
  \bibinfo{author}{\bibfnamefont{Q.}~\bibnamefont{Wu}}, \bibnamefont{and}
  \bibinfo{author}{\bibfnamefont{W.}~\bibnamefont{Yang}}, \bibinfo{journal}{J.
  Chem. Phys.} \textbf{\bibinfo{volume}{123}}, \bibinfo{pages}{062204}
  (\bibinfo{year}{2005}).

\bibitem[{\citenamefont{Cohen et~al.}(2008{\natexlab{a}})\citenamefont{Cohen,
  Mori-S\'{a}nchez, and Yang}}]{Science}
\bibinfo{author}{\bibfnamefont{A.~J.} \bibnamefont{Cohen}},
  \bibinfo{author}{\bibfnamefont{P.}~\bibnamefont{Mori-S\'{a}nchez}},
  \bibnamefont{and} \bibinfo{author}{\bibfnamefont{W.}~\bibnamefont{Yang}},
  \bibinfo{journal}{Science} \textbf{\bibinfo{volume}{321}},
  \bibinfo{pages}{792} (\bibinfo{year}{2008}{\natexlab{a}}).

\bibitem[{\citenamefont{Mori-S\'{a}nchez
  et~al.}(2008)\citenamefont{Mori-S\'{a}nchez, Cohen, and
  Yang}}]{Delocalization}
\bibinfo{author}{\bibfnamefont{P.}~\bibnamefont{Mori-S\'{a}nchez}},
  \bibinfo{author}{\bibfnamefont{A.~J.} \bibnamefont{Cohen}}, \bibnamefont{and}
  \bibinfo{author}{\bibfnamefont{W.}~\bibnamefont{Yang}},
  \bibinfo{journal}{Phys. Rev. Lett.} \textbf{\bibinfo{volume}{100}},
  \bibinfo{pages}{146401} (\bibinfo{year}{2008}).

\bibitem[{\citenamefont{Cohen et~al.}(2008{\natexlab{b}})\citenamefont{Cohen,
  Mori-S\'{a}nchez, and Yang}}]{Fracspin}
\bibinfo{author}{\bibfnamefont{A.~J.} \bibnamefont{Cohen}},
  \bibinfo{author}{\bibfnamefont{P.}~\bibnamefont{Mori-S\'{a}nchez}},
  \bibnamefont{and} \bibinfo{author}{\bibfnamefont{W.}~\bibnamefont{Yang}},
  \bibinfo{journal}{J. Chem. Phys.} \textbf{\bibinfo{volume}{129}},
  \bibinfo{pages}{121104} (\bibinfo{year}{2008}{\natexlab{b}}).

\bibitem[{\citenamefont{Mori-S\'anchez
  et~al.}(2009)\citenamefont{Mori-S\'anchez, Cohen, and Yang}}]{Flatplane}
\bibinfo{author}{\bibfnamefont{P.}~\bibnamefont{Mori-S\'anchez}},
  \bibinfo{author}{\bibfnamefont{A.~J.} \bibnamefont{Cohen}}, \bibnamefont{and}
  \bibinfo{author}{\bibfnamefont{W.}~\bibnamefont{Yang}},
  \bibinfo{journal}{Phys. Rev. Lett.} \textbf{\bibinfo{volume}{102}},
  \bibinfo{pages}{066403} (\bibinfo{year}{2009}).

\bibitem[{\citenamefont{Cohen et~al.}(2009)\citenamefont{Cohen, Mori-S\'anchez,
  and Yang}}]{MP2frac}
\bibinfo{author}{\bibfnamefont{A.~J.} \bibnamefont{Cohen}},
  \bibinfo{author}{\bibfnamefont{P.}~\bibnamefont{Mori-S\'anchez}},
  \bibnamefont{and} \bibinfo{author}{\bibfnamefont{W.}~\bibnamefont{Yang}},
  \bibinfo{journal}{J. Chem. Theory Comput.}  (\bibinfo{year}{2009}),
  \bibinfo{note}{in press.}

\bibitem[{\citenamefont{Yang et~al.}(2009)\citenamefont{Yang, Mori-S\'anchez,
  and Cohen}}]{rpaweitao}
\bibinfo{author}{\bibfnamefont{W.}~\bibnamefont{Yang}},
  \bibinfo{author}{\bibfnamefont{P.}~\bibnamefont{Mori-S\'anchez}},
  \bibnamefont{and} \bibinfo{author}{\bibfnamefont{A.~J.} \bibnamefont{Cohen}}
  (\bibinfo{year}{2009}), \bibinfo{note}{in preparation}.

\bibitem[{\citenamefont{Becke}(1993)}]{B3LYP}
\bibinfo{author}{\bibfnamefont{A.~D.} \bibnamefont{Becke}},
  \bibinfo{journal}{J. Chem. Phys.} \textbf{\bibinfo{volume}{98}},
  \bibinfo{pages}{5648} (\bibinfo{year}{1993}).

\bibitem[{\citenamefont{Iikura et~al.}(2001)\citenamefont{Iikura, Tsuneda,
  Yanai, and Hirao}}]{HiraoLR}
\bibinfo{author}{\bibfnamefont{H.}~\bibnamefont{Iikura}},
  \bibinfo{author}{\bibfnamefont{T.}~\bibnamefont{Tsuneda}},
  \bibinfo{author}{\bibfnamefont{T.}~\bibnamefont{Yanai}}, \bibnamefont{and}
  \bibinfo{author}{\bibfnamefont{K.}~\bibnamefont{Hirao}}, \bibinfo{journal}{J.
  Chem. Phys.} \textbf{\bibinfo{volume}{115}}, \bibinfo{pages}{3540}
  (\bibinfo{year}{2001}).

\end{thebibliography}

\end{document}